\documentclass[twocolumn,amsmath,amssymb,aps,pra,superscriptaddress,showpacs,floatfix,10pt]{revtex4-2}

\usepackage{lmodern}
\usepackage{graphicx}
\usepackage{grffile} 
\usepackage{array}
\usepackage{bbm}
\usepackage{color}
\usepackage[utf8]{inputenc}
\usepackage[T1]{fontenc}
\usepackage{shellesc}
\usepackage{stringenc}

\usepackage{units} 

\usepackage{upgreek} 
\usepackage{slashed}

\usepackage{newtxtext} 
\usepackage{newtxmath}

\usepackage{microtype}

\usepackage{dsfont}
\usepackage{xcolor}

 \newcommand{\bra}[1]{\left\langle{#1}\right|}
 \newcommand{\ket}[1]{\left|{#1}\right\rangle}

 \newcommand{\trace}[1]{\operatorname{Tr}\left\{ #1 \right\}}
 
\newcommand{\I}{i}

\newcommand{\D}{\text{d}}

\def\ba#1\ea{\begin{align}#1\end{align}}																

\usepackage{braket}

\usepackage{tikz}                       
\usepackage{pgfplots}                  
\usetikzlibrary{spy}
\usetikzlibrary{arrows.meta,angles, quotes}

\usetikzlibrary{shapes,arrows,shadows,3d,calc}
\usetikzlibrary{decorations.pathmorphing,decorations.pathreplacing,decorations.markings,trees}
\usetikzlibrary{pgfplots.groupplots}
\usepackage{tikz-3dplot}

\pgfplotsset{%
  every axis legend/.append style={%
    cells={anchor=west},
    at={(0.96,0.04)},
    anchor=south east,
    font=\scriptsize
  },
  every axis/.append style={%
    yticklabel style={/pgf/number format/fixed zerofill, /pgf/number format/precision=2}
  },
  width= 0.45\textwidth, height=5cm, xmajorgrids=false, xminorgrids=false, minor x tick num=1
}

\usepackage[hang]{subfigure} 

\pgfplotsset{compat=1.18}

\usepackage{blindtext} 
\usepackage{booktabs}

\newcommand*{\balancecolsandclearpage}{%
  \close@column@grid
  \cleardoublepage
  \twocolumngrid
}

\definecolor{cset-aps-blueberry}{RGB}{28,128,158}
\definecolor{cset-aps-blue}{RGB}{46,44,184}
\definecolor{cset-aps-turquoise}{RGB}{0,67,88}
\definecolor{cset-aps-limegreen}{RGB}{190,219,67}
\definecolor{cset-aps-green}{RGB}{31,138,112}
\definecolor{cset-aps-yellow}{RGB}{255,225,25}
\definecolor{cset-aps-orange}{RGB}{253,116,0}
\definecolor{cset-aps-red}{RGB}{219,0,43}
\usepackage{hyperref}
\hypersetup{%
	colorlinks=true,
	linkcolor={cset-aps-red},
	linkbordercolor={cset-aps-red},
	filecolor={cset-aps-orange},
	filebordercolor={cset-aps-orange},
	citecolor={cset-aps-blue},
	citebordercolor={cset-aps-blue},
	urlcolor={cset-aps-green},
	urlbordercolor={cset-aps-green},
	menucolor={cset-aps-limegreen},
	menubordercolor={cset-aps-limegreen},
	breaklinks=true,
	pdfborderstyle={/S/U/W 2},
	pdfpagemode=UseOutlines,
	pdfstartpage={1},
}

\begin{document}
\title{Entangled photons from para-positronium decay: Do coincidences from scattered photons imply a Bell state?}
\author{Paul Joos}
\affiliation{German Aerospace Center (DLR), Institute of Quantum Technologies, Wilhelm-Runge-Stra{\ss}e 10, D-89081 Ulm, Germany}
\author{Peter Kling}
\affiliation{German Aerospace Center (DLR), Institute of Quantum Technologies, Wilhelm-Runge-Stra{\ss}e 10, D-89081 Ulm, Germany}


\begin{abstract}
Electron and positron can form a meta-stable bound state called positronium that decays via pair annihilation. We show how polarization-dependent Compton scattering can be used to verify that the two annihilation photons in the spin-zero case (para-positronium) are emitted in a maximally entangled Bell state. Our theoretical  approach based on two-photon density matrices connects concepts from relativistic quantum electrodynamics and quantum information theory.    
\end{abstract}

\maketitle

\section{Introduction}

According to the results of quantum electrodynamics, the photons emitted by the para-positronium (para-Ps) decay are maximally entangled with respect to their polarizations~\cite{bohm1957,latorre2001}. So far, there exists no verification of this entanglement, that does not come along with additional assumptions~\cite{bohm1957}. For the high energies of the annihilation photons ($511\,\text{keV}$) standard polarimeters cannot be used and one has to resort to the polarization dependence of Compton scattering. In this paper we show the following statement: The predicted angular distribution~\cite{pryce1947,snyder1948} for coincidences of scattered photons from the para-Ps decay is obtained,  \emph{if and only if} the emitted photons have been in a maximally entangled Bell state prior scattering.   

This proof is enabled by our representation of this distribution, that factorizes into a contribution containing information on the two-photon density matrix and a part corresponding to the geometry of the detection setup. Moreover, we sketch a scheme for efficiently tracing back to the Bell state without scanning the complete solid angles, that relies on the comparison of the number of counts for different geometrical configurations slightly modifying the usual approaches~\cite{bohm1957}. Lastly, we study effects of finite angular widths reflecting the nonzero resolutions of the detectors.       

John Wheeler already developed in 1946 the idea of detecting coincidences of scattered photons from the para-Ps decay~\cite{wheeler1946}. Shortly thereafter, the correct angular distribution for such coincidences was calculated predicting the maximum value $2.84$ for the ratio of counts detected (i) perpendicularly to each other and  (ii) parallel~\cite{pryce1947,snyder1948}. After the first experimental realizations~\cite{bleuler1948,hanna1948}, the experiment of Wu and Shaknov in 1950 demonstrated for the first time a reliable consistency with theory~\cite{wu1950}.         

In 1957 Bohm and Aharanov made the connection of the Wu \& Shaknov experiment to the EPR paradox: The entangled state from quantum theory was compared with a separable one and it was shown that the experimental results strongly imply the entanglement of the photons~\cite{bohm1957}. However, as discussed in Refs.~\cite{clauser1969,horne1970}, the polarization measurement via Compton scattering does not enable a violation of Bell inequalities as a sufficient condition to rule out hidden variable theories. It was argued that a mapping of the discrete degrees of freedom of polarization to partitions of the scattering sphere impedes such a violation~\cite{horne1970}. Hence, the focus of research successfully shifted to the optical regime, where suitable polarizers are available~\cite{genovese2005}. Nevertheless, several experiments regarding the coincidences of scattered annihilation photons were performed during the years  ~\cite{bertolini1955,langhoff1960,kasday1972,faraci1974,kasday1975,wilson1976,bohm1976,bruno1977,bertolini1981,osuch1996}.

The topic gained renewed interest as proposals to use the polarization correlation for improving medical imaging techniques, in particular positron emission tomography (PET), emerged~\cite{kuncic2011,mcnamara2014,toghyani2016}. In the following years, several experiments~\cite{makek2020,watts2021,ivashkin2023,bordes2024,parashari2024,moskal2025,tkachev2025} were performed, for example with the Jagiellonian PET (J-PET) device~\cite{moskal2018}. In parallel,  much theoretical effort was invested, for example in Refs.~\cite{hiesmayr2019,hiesmayr2024,bala2025,funck2025,zugec2026,clarke2026} or in a series of papers~\cite{caradonna2019,caradonna2024,caradonna2024_annals,caradonna2025,caradonna2025entanglement,caradonna2026} that employed a Stokes vector formulation~\cite{fano1949,mcmaster1961} to describe the photon polarization.

We complement these approaches with a formalism~\cite{wightman1948} that is  based directly on the density matrix. This procedure helps us to study entanglement and facilitates the connection of concepts from relativistic quantum electrodynamics and quantum information theory.   

This paper is organized as follows: We begin in Sec.~\ref{sec:Photon_state} by writing down the two-photon state of the annihilation photons according to quantum electrodynamics. In Sec.~\ref{sec:Polarization-dependent_Compton_scattering} we discuss the basic setup for detecting coincidences from scattered photons and we explain how measuring the predicted angular distribution implies that the photons were indeed in a Bell state before scattering. Finally, we discuss our main results and conclude in Sec.~\ref{sec:Discussion}. To keep this paper self-contained we add Appendices~\ref{sec:Para-positronium_decay} and~\ref{sec:Compton_scattering}, where we present the detailed calculations for obtaining the two-photon state for the para-Ps decay and the angular distribution for Compton scattering, respectively, both  starting from first principles.

\section{Photon state}
\label{sec:Photon_state}

Ps represents a meta-stable bound state of the electron and its anti-particle the positron~\cite{berko1980,rich1981,adkins2022,bass2023}. Depending on the spin configuration of its constituents Ps can either form a singlet (para-Ps) with total spin zero or a triplet (ortho-Ps) with total spin one. In the present article we only consider ground-state para-Ps which decays in leading order into two photons via pair annihilation~\cite{greiner}.

The results from quantum electrodynamics show that the emitted photons are entangled in polarization~\cite{jauch,latorre2001}. In App.~\ref{sec:Para-positronium_decay} we recall the calculations from Ref.~\cite{latorre2001} and obtain the two-photon state
\begin{equation}
    \ket{\psi_\text{p-Ps}} \propto
    \int\!\!\D^3k\,\updelta(|\boldsymbol{k}|-m) \ket{\psi_\text{p-Ps}(\boldsymbol{k})}
    \label{eq:psi_para}
\end{equation}
with
\begin{equation}
\begin{aligned}
 \ket{\psi_\text{p-Ps}(\boldsymbol{k})}&=  \frac{1}{\sqrt{2}} \Big(\ket{+(\boldsymbol{k}),+(-\boldsymbol{k})}-\ket{-(\boldsymbol{k}),-(-\boldsymbol{k})}\Bigr)\\
 &=\frac{1}{\sqrt{2}}\Bigl(\ket{1(\boldsymbol{k}),2(\boldsymbol{-k})}+\ket{2(\boldsymbol{k}),1(-\boldsymbol{k})}\Bigr)
\end{aligned}
 \label{eq:psi_k_para}
\end{equation}
 after the para-Ps decay. For convenience, we have chosen the units such that $\hbar=c=1$. The photon pairs are emitted with equal probability in any direction. However, the photons within one distinct pair have always opposite direction due to momentum conservation, that is $\boldsymbol{k}_\text{A}=-\boldsymbol{k}_\text{B}\equiv \boldsymbol{k}$. Energy conservation implies that the energy of each photon equals the rest mass of an electron, that is $|\boldsymbol{k}|\equiv k=m $, which is ensured by the delta function in Eq.~\eqref{eq:psi_para}. We have written the state in Eq.~\eqref{eq:psi_k_para} in terms of circular ($+,-$) and linear polarization ($1,2$), respectively, which are connected via the relation $\displaystyle\ket{\pm}=2^{-1/2}(\ket{1}\pm i\ket{2})$. In both bases the state  represents a maximally entangled Bell state: $\ket{\Phi_-}$ for circular and $\ket{\Psi_+}$ for linear polarization~\cite{caradonna2024}. For the remainder of the article we consider linear polarization. We note that the form of Eq.~\eqref{eq:psi_k_para} also follows from selection rules imposed by fundamental symmetries~\cite{yang1950,jauch,caradonna2024}.

\section{Polarization-dependent Compton scattering}
\label{sec:Polarization-dependent_Compton_scattering}
\begin{figure}
    \centering
    \includegraphics[width=1\columnwidth]{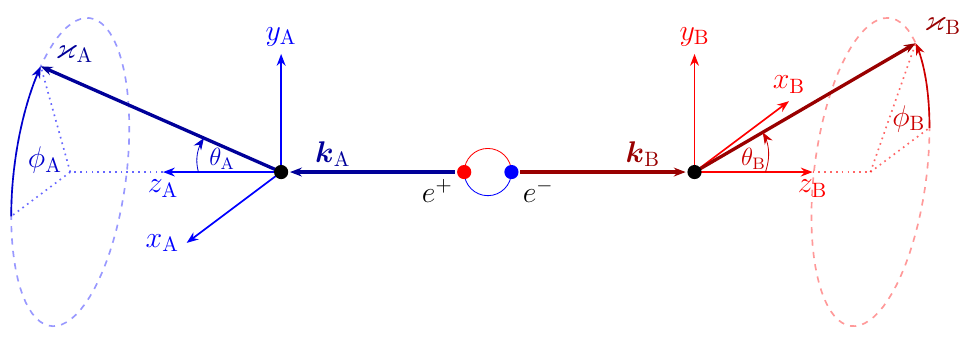}

    \caption{Setup for determining two-photon coincidences from para-Ps decay and subsequent polarization-dependent Compton scattering: The photons are emitted at the origin in opposite directions with two scatterers equidistantly placed on both sides in a line with the source. The coordinate systems for the two observers, A and B, are defined such that the $z$-axes are in the direction of the respective photon, that is $z_\text{A}=-z_\text{B}$, and we further choose $x_\text{A}=-x_\text{A}$ and $y_\text{A}=y_\text{B}$. On each side detectors are placed such that we can measure coincidences as a function of the scattering angles $\theta_\text{A}$, $\theta_\text{B}$ and azimuths $\phi_\text{A}$, $\phi_\text{B}$  (compare to Eq.~\eqref{eq:n_k} for the relation of the propagation direction and the corresponding angles).}
    \label{fig:setup}
\end{figure}

Our setup is depicted in Fig.~\ref{fig:setup}: At the origin, photons from the Ps decay are emitted. Two scatterers are equidistantly placed from the source such that the source and the scatterers form a straight line. Hence, only photons directed along this line are scattered reducing the isotropic state in Eq.~\eqref{eq:psi_para} to a single wave-vector component in Eq.~\eqref{eq:psi_k_para}. To further decrease the influence of unwanted directions we could coat the source with a lead sphere with a small canal through the origin and place the scatterers directly behind the openings of this channel as proposed in the original paper of Wheeler~\cite{wheeler1946} and implemented for example in Ref.~\cite{wu1950}. Alternatively, an inner ring of detectors can simultaneously act as scatterers such that it is always clear which path the photons took to end up at the outer and final ring of detectors~\cite{moskal2018}.   

\subsection{Two-photon coincidences}
\label{sec:two_photon_coincidences}

We now detect the scattered photons as functions of the scattering angle $\theta$ and the azimuth $\phi$ (compare to Fig.~\ref{fig:setup}). By installing detectors on both sides we can measure coincidences.
This setup resembles schemes for quantum state tomography of optical photon pairs~\cite{james2001} with the difference that we have to resort to Compton scattering instead of employing standard polarizers.    

In App.~\ref{sec:Compton_scattering} we derive the probability distribution
\begin{equation}
 \label{eq:p}
    p(\theta_\text{A},\phi_\text{A},\theta_\text{B},\phi_\text{B}) =\mathcal{N} \,\text{vec}{(W^\text{T})}^\text{T} \cdot v(\theta_\text{A},\phi_\text{A})\otimes v(\theta_\text{B},\phi_\text{B})
\end{equation}
for coincidences with an arbitrary two-photon density matrix $\rho$ in (linear) polarization representation. This distribution depends on the scattering angles $\theta_\text{A},\theta_\text{B}$ and azimuthal angles $\phi_\text{A},\phi_\text{B}$. We note that the expression in Eq.~\eqref{eq:p} can be separated into (i) a term containing information on the initial density matrix $\rho$ and (ii) a geometric contribution corresponding to the setup.    

We express the latter part with the help of the column vector   
\begin{equation}
    v(\theta,\phi)\equiv\left(\frac{\varkappa(\theta)}{k}\right)^2
    \begin{pmatrix}
    S(\theta)\\ - f(\theta) \cos{2\phi} \\ -f(\theta)\sin{2\phi}
    \end{pmatrix}
    \label{eq:v}
\end{equation}
and the $\theta$-dependent functions
\begin{align}
\label{eq:S_theta}
    S(\theta)&= \frac{\varkappa(\theta)}{k}+\frac{k}{\varkappa(\theta)}-\sin^2{\theta}\,,\\
\label{eq:f_theta}
    f(\theta)&=\sin^2{\theta}\,,\\
\label{eq:kappa_theta}
    \varkappa(\theta)&=\frac{k}{1+\frac{k}{m}(1-\cos{\theta})}
\end{align}
in the notation of Ref.~\cite{horne1970}. The wave number $\varkappa=\varkappa(\theta)$ of  a scattered photon depends via the Compton formula, Eq.~\eqref{eq:kappa_theta}, on the scattering angle and on the initial wave number $k$. For the sake of completeness, we quote the constant~\cite{snyder1948}
\begin{equation}
    \mathcal{N}=\frac{1}{\left[2\pi\left(\frac{40}{9} -3\log{3} \right)\right]^2}
\end{equation}
normalizing the distribution to unity with respect to the solid angles on both sides, where we have set $k=m$ for the para-Ps decay.  

The contribution corresponding to the two-photon density matrix prior scattering is determined by the matrix $W$ whose elements are formed via the prescription 
\begin{equation}
    W_{ij}\equiv \trace{u_i\otimes u_j\cdot \rho}
    \label{eq:W}
\end{equation}
with the help of the column vector
\begin{equation}
    u\equiv
    \begin{pmatrix}
    \mathds{1}\\\sigma_z\\\sigma_x
    \end{pmatrix}
    \label{eq:u}
\end{equation}
containing the identity matrix and two of the three Pauli matrices. We note that the vectorization operation $\text{vec}(A)$ in Eq.~\eqref{eq:p} stacks the three columns of a $3\times3$ matrix $A$  such that a column vector with nine entries emerges. 

In the following we derive two statements: Firstly, we calculate the coincidence distribution of scattered photons from $\ket{\Psi_+}$ which is straightforward due to Eq.~\eqref{eq:p}. Secondly, we show that this result emerges \textit{only}, if the initial state is given by this Bell state.

The density matrix corresponding to $\ket{\Psi_+}$ reads
\begin{equation}
    \rho_\text{p-Ps}=\ket{\psi_{\text{p-Ps}}(\boldsymbol{k})}\bra{\psi_{\text{p-Ps}}(\boldsymbol{k})}=\frac{1}{2}
    \begin{pmatrix}
    0 & 0 & 0 & 0\\
    0 & 1 & 1 & 0\\
    0 & 1 & 1 & 0\\
    0 & 0 & 0 & 0
    \end{pmatrix}
    \label{eq:rho_psi_+}
\end{equation}
in (linear) polarization representation, where we have omitted the dependency on the wave vector $\boldsymbol{k}$. We find the distribution~\cite{pryce1947,snyder1948}
\begin{equation}
 \label{eq:p_Ps}
     \begin{aligned}
     p_{\text{p-Ps}}&(\theta_\text{A},\phi_\text{A},\theta_\text{B},\phi_\text{B})=\mathcal{N} 
    \left(\frac{\varkappa(\theta_\text{A})}{m}\right)^2\left(\frac{\varkappa(\theta_\text{B})}{m}\right)^2\\
    &\times\Bigl\{S(\theta_\text{A})S(\theta_\text{B})-f(\theta_\text{A})f(\theta_\text{B})\cos{[2(\phi_\text{A}+\phi_\text{B})]}\Bigr\}
 \end{aligned}
\end{equation}
for the coincidences resulting from Eq.~\eqref{eq:p} with $\rho=\rho_\text{p-Ps}$ from Eq.~\eqref{eq:rho_psi_+}.  

Conversely, by inserting an \emph{arbitrary} $4\times 4$ density matrix into Eq.~\eqref{eq:p} we derive  the conditions $W_{11}=W_{33}=-W_{22}=1$ for the diagonal elements of $W$ while $W_{ij}=0$ for $i\neq j$, if we assume that Eq.~\eqref{eq:p_Ps} holds and equate the coefficients. These relations translate to the following constraints
\begin{align}
    \label{eq:Tr1}
    \trace{\rho}=\trace{\sigma_x\otimes \sigma_x \cdot \rho}&=-\trace{\sigma_z\otimes \sigma_z \cdot \rho}=1\\
    \label{eq:Tr2}
    \trace{u_i\otimes u_j \cdot \rho} &=0\ \ \ \ i\neq j
\end{align}
for the density matrix. Of course, the trace of a density operator should always equal unity to ensure the correct normalization. Moreover, $\rho$ has to be  Hermitian and positive semi-definite to describe meaningful probabilities~\cite{nielsen}. 

With the help of Eqs.~\eqref{eq:Tr1} and~\eqref{eq:Tr2} and by demanding Hermiticity we obtain the expression
\begin{equation}
 \rho=    
    \begin{pmatrix}
     0 & \I a & \I b & \frac{1}{2}-\alpha +\I\gamma\\
     -\I a & \frac{1}{2}&\alpha+\I\beta & \I c\\
     -\I b& \alpha-\I\beta &  \frac{1}{2} & \I d\\
     \frac{1}{2}-\alpha - \I\gamma & -\I c & -\I d & 0 
    \end{pmatrix}
\end{equation}
for the density matrix containing seven real parameters $a,b,c,d,\alpha,\beta,\gamma$. We observe that two diagonal elements of $\rho$ are zero. In this case, all elements of the corresponding rows and columns have to be zero as well, which is a necessary condition to ensure positive semi-definiteness~\cite{horn}. We can motivate this mathematical statement by realizing that if there are no populations (diagonal elements) of the states $\ket{1,1}$ and $\ket{2,2,}$, there also cannot be coherences (off-diagonal elements) for these states. Hence, we 
deduce that $a=b=c=d=\gamma=0$ while $\displaystyle\alpha=1/2$. It is then straightforward to calculate the (possibly) nonzero eigenvalues $\displaystyle\lambda_{\pm}=\frac{1}{2}\pm \frac{1}{2} \sqrt{1+4\beta^2}$ of $\rho$. Since $\lambda_-< 0$ for $|\beta|>0$ we conclude that $\beta=0$ since a negative eigenvalue would contradict the semi-definiteness of $\rho$. As a result, we are left with the expression for $\rho$ in Eq.~\eqref{eq:rho_psi_+}, that is, the Bell state $\ket{\Psi_+}$ is the only possible initial condition for which the distribution in Eq.~\eqref{eq:p_Ps} can be observed. In other words: If we can measure  $p(\theta_\text{A},\phi_\text{A},\theta_\text{B},\phi_\text{B})$ completely and observe the result in Eq.~\eqref{eq:p_Ps},  the annihilation photons must have been in the Bell state $\ket{\Psi_+}$ before scattering.   

\subsection{Ratios for different measurement angles}

\begin{figure}
    \centering
    \includegraphics[width=\columnwidth]{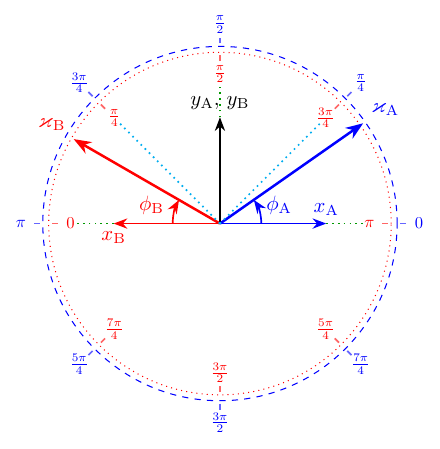}
    \caption{Top view of the $x-y$ plane of our setup: We illustrate how the values of the azimuths  $\phi_\text{A}$ (blue) and  $\phi_\text{B}$ (red)  correspond to each other. For example, the ratio in Eq.~\eqref{eq:r} compares cases, where the
    counts are detected perpendicular to each other ($\phi_\text{A}=0$ and $\phi_\text{B}=\pi/2$) and where they are anti-parallel ($\phi_\text{A}=0$ and $\phi_\text{B}=0$). Similarly, the events considered in Eq.~\eqref{eq:r_pi4} are either perpendicular ($\phi_\text{A}=\pi/4$ and $\phi_\text{B}=\pi/4$) or parallel ($\phi_\text{A}=\pi/4$ and $\phi_\text{B}=3\pi/4$).  }
    \label{fig:ratios}
\end{figure}

 In  experiments~\cite{wu1950,moskal2025} often ratios of counts for two different geometric configurations were determined. This procedure eliminates the need for scanning the entire solid angle. These ratios depend on the quantity
\begin{equation}
    a(\theta)\equiv \frac{f(\theta)}{S(\theta)}\leq 0.69
    \label{eq:a}
\end{equation}
which attains its maximum at $\theta_0\cong 81.7^\circ$~\cite{pryce1947}. We fix both scattering angles to this value, that is $\theta_\text{A}=\theta_\text{B}=\theta_0$, while comparing the results for different values of the azimuths. For example, we consider the ratio     
\begin{equation}
    \frac{p(\theta_0 | 0, \pi/2)}{p(\theta_0|0,0)}=
    \frac{1+a^2(\theta_0)}{1-a^2(\theta_0)}\equiv R(\theta_0)\cong 2.84
    \label{eq:r}
\end{equation}
of the number of counts in perpendicular directions to the number of counts in anti-parallel directions (compare to Fig.~\ref{fig:ratios})~\cite{wheeler1946,pryce1947,snyder1948,wu1950}. Indeed, this measurement alone cannot distinguish an entangled -- or even Bell -- state from a separable one, unless reasonable restrictions to a hypothetical separable state are applied (see App.~\ref{sec:Compton_scattering})~\cite{bohm1957,caradonna2024}. 

However, we can add modifications to this setup to directly extract the initial Bell state. Instead of measuring solely coincidences we can detect counts on only one side while disregarding the other side and vice versa. The reduced density operators of $\rho_\text{p-Ps}$ in Eq.~\eqref{eq:rho_psi_+} describe a completely unpolarized state for each side, that is  $\rho_\text{A}=\rho_\text{B}=\frac{1}{2}\mathds{1}$. From the Klein--Nishina formula (see App.~\ref{sec:Compton_scattering}) we obtain the probabilities  $p_\text{A}(\theta_\text{A},\phi_\text{A})\propto S(\theta_\text{A})$
and
$p_\text{B}(\theta_\text{B},\phi_\text{B})\propto S(\theta_\text{B})$
that are independent of $\phi_\text{A}$ and $\phi_\text{B}$. By placing two detectors at the same side of the experiment but at different azimuth angles we find that the ratios
\begin{equation}
\label{eq:ra_rb}
    \frac{p_\text{A}(\theta_0|\pi/2)}{p_\text{A}(\theta_0|0)}=\frac{p_\text{B}(\theta_0|\pi/2)}{p_\text{B}(\theta_0|0)}=1
\end{equation}
equal unity reflecting the rotational symmetry with respect to the $z$-axis. For Eq.~\eqref{eq:ra_rb} to hold the density matrix has to satisfy the relations 
\begin{equation}
\label{eq:Tr3}
    \trace{\sigma_z\otimes\mathds{1}\cdot \rho}=
    \trace{\mathds{1}\otimes \sigma_z \cdot \rho}=0\,.
\end{equation}
Moreover, Eq.~\eqref{eq:r} requires that  
\begin{equation}
\label{eq:Tr4}
    \trace{\sigma_z \otimes \sigma_z \cdot \rho}=-1
\end{equation}
is fulfilled.

With the help of Eqs.~\eqref{eq:Tr3}  and~\eqref{eq:Tr4} 
together with the condition $\trace{\rho}=1$, we obtain the expression
\begin{equation}
    \rho= 
    \begin{pmatrix}
    0 & 0 & 0 & 0 \\
    0 & 1/2 & z & 0 \\
    0 & z^*& 1/2 & 0 \\
    0 & 0 & 0 & 0
    \end{pmatrix}
    \label{eq:rho_z}
\end{equation}
for the density matrix, where we have already used that all entries in rows and columns corresponding to zero diagonal elements have to be zero as well for positive semi-definite matrices (analogously to the preceding section)~\cite{horn}. The remaining complex parameter $z$ has to fulfill the condition $|z|\leq 1/2$ ensuring that all eigenvalues of $\rho$ are non-negative.
For nonzero $z$ the application of the Peres–Horodecki criterion~\cite{peres1996,horodecki2001} shows that $\rho$ is non-separable since at least one eigenvalue of the partial transpose of $\rho$ is negative. However, the case $z=0$ corresponds to the separable state $\displaystyle \rho=\frac{1}{2}\left(\ket{1,2}\bra{1,2}+\ket{2,1}\bra{2,1}\right)$
~\footnote{We note that this state does not show a rotational symmetry with respect to the $z$-axis as premised in Refs.~\cite{bohm1957,caradonna2024}.}. Hence, we require additional measurements to verify entanglement.

Indeed, the further ratio (compare to Fig.~\ref{fig:ratios})
\begin{equation}
\label{eq:r_pi4}
    \frac{p(\theta_0 | \pi/4, \pi/4)}{p(\theta_0|\pi/4,3\pi/4)}=
    R(\theta_0)
\end{equation}
of coincidences for the angles $\phi_\text{A}=\pi/4$, $\phi_\text{B}=\pi/4$ (perpendicular to each other) and $\phi_\text{A}=\pi/4$, $\phi_\text{B}=3\pi/4$ (parallel) also yields the maximal value of $2.84$ for the Bell state $\ket{\Psi_+}$. However, in this case, we do not require Eq.~\eqref{eq:Tr4} to hold, but instead we find that
\begin{equation}
    \trace{\sigma_x \otimes \sigma_x \cdot \rho}=1
\end{equation}
is a necessary condition. For the density matrix in Eq.~\eqref{eq:rho_z} this constraint translates to 
$\text{Re}(z)=1/2$.   
Together with $|z|\leq 1/2$ we arrive at $z=1/2$, that is, the density matrix is of the form of Eq.~\eqref{eq:rho_psi_+} and describes the Bell state $\ket{\Psi_+}$. 

Hence, if  we (i) measure the value 2.84 for, both,  the ratios of coincidences  in Eqs.~\eqref{eq:r}  and~\eqref{eq:r_pi4} and (ii) observe the symmetry with respect to the $z$-axis for each side separately with the help of the ratios in Eq.~\eqref{eq:ra_rb}, we know that the photons were in a Bell state before scattering.     

\subsection{Finite angle intervals}

So far, we have assumed a definite value for the measured angles, which represents a simplification since every real detector has a finite resolution. Already in Ref.~\cite{snyder1948} this issue was recognized and the effect of finite angle intervals was calculated.
\pgfplotsset{
    colormap={mathematica_muted}{
        rgb255=(70, 100, 170)   
        rgb255=(180, 110, 140)  
        rgb255=(245, 235, 180)  
    }
}
\begin{figure}
    \centering
     \begin{tikzpicture}
        \begin{axis}[
        width=0.85\columnwidth, 
        height=0.75\columnwidth,
        xlabel={$\Delta\theta$},
        ylabel={$\Delta\phi$},
        xtick={0,pi/4,1.42733},
        xticklabels={0,$\displaystyle\frac{\pi}{4}$,$\displaystyle\theta_0$},
        ytick={0,pi/4,pi/2},
        yticklabels={0,$\displaystyle\frac{\pi}{4}$,$\displaystyle\frac{\pi}{2}$},
        xmin=0.0001, xmax=1.42733,   
        ymin=0, ymax=1.5708,
        colorbar, 
        colorbar style={title={$R(\theta_0,\Delta\theta,\Delta\phi)$},
        ylabel style={font=\footnotesize},
        ytick={1,1.5,2,2.5,3},
        yticklabels={1.0,1.5,2.0,2.5,3.0}},
        colormap name=mathematica_muted,
        point meta min=1, 
        point meta max=3,  
        enlargelimits=false, 
        axis on top
    ]
     \addplot graphics [
            xmin=0.0001, xmax=1.42733, 
            ymin=0, ymax=1.5708
        ] {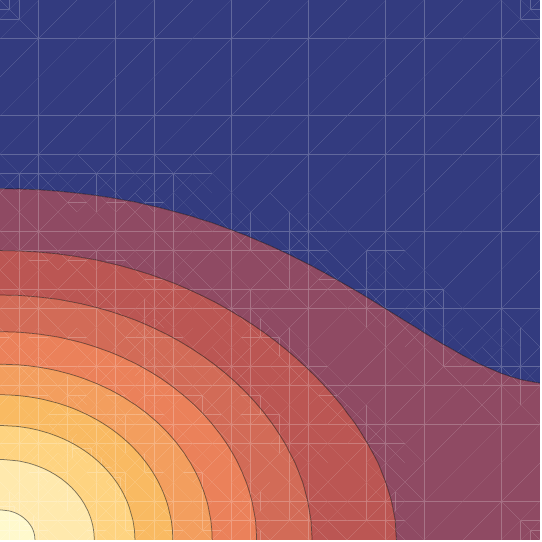};
    
    \end{axis}
     \end{tikzpicture}
    \caption{Contour plot of the ratio $R$ of perpendicular to (anti-)parallel counts as a function
    of the angular widths $\Delta\theta$ and $\Delta\phi$. We note that $R=R(\theta_0,\Delta\theta,\Delta\phi)$ is a generalization of the expression in Eq.~\eqref{eq:r} with the substitution of $a(\theta_0)$ from Eq.~\eqref{eq:a} by $a(\theta_0, \Delta\theta, \Delta\phi)$ from Eq.~\eqref{eq:a_delta}. In all cases we set the mean value of the scattering angle to $\bar{\theta}_\text{A}=\bar{\theta}_\text{B}=\theta_0\cong 81.7^\circ$ while we compare the number of counts for two different mean azimuths $\bar{\phi}_\text{A}$ and $\bar{\phi}_\text{B}$ analogously to Eq.~\eqref{eq:r} or Eq.~\eqref{eq:r_pi4}. We observe that for larger angular widths, with $\bar{\theta}_i\pm \Delta \theta$  and $\bar{\phi}_i\pm \Delta \phi$, the value of $R$ decreases from its maximum value $R(\theta_0)\cong 2.84 $ at $\Delta\theta,\Delta\phi \rightarrow 0$, as already derived in Ref.~\cite{snyder1948}, with a slightly slower decrease in the $\Delta\phi$-direction. Ultimately $R$ approaches unity  for increasing $\Delta\theta,\Delta\phi$ making it more and more difficult to distinguish the result of the Bell state $\ket{\Psi_+}$ from any other -- eventually separable -- state showing the same behavior.}
    \label{fig:breiten}
\end{figure}

We consider the symmetric intervals $\bar{\theta}_i\pm \Delta \theta$ and $\bar{\phi}_i\pm \Delta \phi$. Similarly to the preceding section we set $\bar{\theta}_i=\theta_0$ for both sides while varying $\bar{\phi}_\text{A}$ and $\bar{\phi}_\text{B}$. The resulting distribution of coincidences
reads
\begin{equation}
 \begin{aligned}
    p(\theta_0;& \Delta \theta |\bar{\phi}_\text{A},\bar{\phi}_\text{B};\Delta \phi) =\mathcal{N} \,\text{vec}{(W^\text{T})}^\text{T}
    \\
    &\times v(\theta_0,\Delta \theta, \bar{\phi}_\text{A},\Delta \phi)\otimes  v(\theta_0,\Delta \theta, \bar{\phi}_\text{B},\Delta \phi)\,,
 \end{aligned}
\end{equation}
where we have introduced 
\begin{equation}
    v(\theta_0,\Delta \theta, \bar{\phi},\Delta \phi)\equiv \frac{1}{\Delta \Omega}
    \int\limits_{\bar{\phi}-\Delta\phi}^{\bar{\phi}+\Delta\phi} \!\!\D\phi \int\limits_{\theta_0-\Delta\theta}^{\theta_0+\Delta\theta} \!\!\D\theta \sin{\theta} \, v(\theta,\phi)
\end{equation}
as a generalization of the expression in Eq.~\eqref{eq:v} with $\Delta \Omega \equiv4\Delta\phi\sin{\theta_0}\sin{\Delta\theta} $ as the interval of the solid angle. The procedure of the preceding section can now easily be adapted to finite angle intervals by substituting the parameter $a_0(\theta)$ with the generalized version  
\begin{equation}
    a(\theta_0, \Delta\theta, \Delta\phi)
    \equiv \text{sinc}{(2\Delta\phi)}\,
    \frac{\int\limits_{\theta_0-\Delta\theta}^{\theta_0+\Delta\theta}\!\!\D\theta \sin{\theta} \, f(\theta)}{\int\limits_{\theta_0-\Delta\theta}^{\theta_0+\Delta\theta}\!\!\D\theta \sin{\theta} \, S(\theta)}
    \label{eq:a_delta}
\end{equation}
depending on the widths $\Delta\theta$ and $\Delta\phi$. In Fig.~\ref{fig:breiten} we have plotted the result for the ratio $R=R(\theta_0,\Delta\theta,\Delta\phi)$ as a function of the angular widths.

If we can precisely determine the mean values and if know the values of $\Delta\theta$ and $\Delta\phi$, we know how much the ratio $R(\theta_0,\Delta\theta,\Delta\phi)$ is reduced from its maximum $2.84$ and we repeat the arguments from the preceding section to verify if the photons are initially in a Bell state. In practice, this verification becomes more difficult for increasing values of $\Delta\theta$ and $\Delta\phi$, because the results for different initial states approach unity and ultimately they cannot be distinguished from another.   

\section{Discussion}
\label{sec:Discussion}
Detection of coincidences from Compton scattered photons makes it possible to verify that the annihilation photon were in a Bell state with respect to their polarization. It suffices to place four detectors one each side at the proposed angles and consider ratios (i) of counts on each side separately and (ii) of coincidences. Hence, our scheme only slightly modifies or reinterprets existing ones in the way that measurement data are recorded and further analyzed. 

So far, the maximum value $R(\theta_0)=2.84$ was never completely verified  in experiments~\cite{moskal2025}. This fact immediately raises the question of the underlying cause. One possible source of inconsistency between theory and experiment is the detection scheme. If the deviation comes from a finite angular resolution this issue can be straightforwardly solved by taking finite angular widths into account by means of Eq.~\eqref{eq:a_delta}~\cite{snyder1948} leading to a decreased value of $R$. If that procedure does not suffice, one has to continue searching for a more accurate modelling of the scheme, for example in a more detailed description of Compton scattering from bound electrons~\cite{sommerfeldt2026} or in the geometry of the setup. 

Another possible source of inconsistency between experiment and theory could of course emerge, if the initial state is not a Bell state. According to Refs.~\cite{moskal2025,kumar2025}, the measured value of $R$ depends on the specific Ps source and the authors concluded that the emission of photons from different processes beyond the para-Ps decay (direct annihilation, pick-off annihilation of ortho-Ps) lead to changed results. In this case, the two-photon density matrix would be described by the mixture $P\rho_\text{p-Ps} +(1- P)\tilde{\rho}$ of the Bell state and an unknown density matrix $\tilde{\rho}$ with a source-dependent probability $P$~\cite{moskal2025}.  This approach could be a promising starting point for future theoretical and experimental efforts.    

A different strategy is to find lower limits of $R$ (or of similar measures) that still verify entanglement. In real-life experiments there will always be a  uncertainty such that $R(\theta_0)=2.84$  could be validated only up to a certain precision. Our derivation in Sec.~\ref{sec:Polarization-dependent_Compton_scattering}, however, strongly relies on the on-off criterion that two diagonal elements of the density matrix are exactly zero. Deviations from the theoretical predictions lead to very small but non-zero diagonal elements and there will be now a whole class of possible density matrices. One can study the properties of this class by means of standard entanglement criteria/measures (Peres–Horodecki criterion, concurrence, Schmidt-number, etc.) or with the help of alternative methods, such as for example in  Refs.~\cite{tschaffon2023,tschaffon2024}.

\begin{acknowledgements}
We thank Maxim~A. Efremov, Michael Tschaffon, Jan Funkler, and Alexander Wolf for many fruitful discussions. 
\end{acknowledgements}

\begin{appendix}
\section{Para-positronium decay}
\label{sec:Para-positronium_decay}

In this appendix, we calculate the two-photon state from the para-Ps decay by means of relativistic quantum electrodynamics. Hereby, we closely follow  the lines of Refs.~\cite{latorre2001,jauch,itzykson,sen2019}.

Prior to interaction, the state vector 
\begin{equation}
  \begin{aligned}
       \ket{\text{Ps}}=\int\!\!\D^3P & \, \Psi(\boldsymbol{P}) \int\!\!\ \D^3p \, \psi(\boldsymbol{p})\, \\
    & \times\sum\limits_{s_1,s_2}\mathcal{R}_{s_1,s_2}^{S,S_z}\,\,
    \hat{b}_{s_1}^\dagger(\boldsymbol{p}_1)\,
    \hat{d}_{s_2}^\dagger(\boldsymbol{p}_2)\ket{0}
  \end{aligned}
  \label{eq:app_Ps}
\end{equation}
of the Ps~\cite{greiner,sen2019} is described by the wave functions, $\Psi=\Psi(\boldsymbol{P})$ and $\psi=\psi(\boldsymbol{p})$, of center-of-mass and relative motion, respectively.
 We note that the creation operators $\hat{b}_{s_1}^\dagger=\hat{b}_{s_1}^\dagger(\boldsymbol{p}_1)$ and $\hat{d}_{s_2}^\dagger=\hat{d}_{s_2}^\dagger(\boldsymbol{p}_2)$ of electron and positron depend on the momenta $\boldsymbol{p}_1$ and $\boldsymbol{p}_2$ of the single particles. In our formalism, these momenta are functions of the center-of mass momentum and the relative momentum, that is $\boldsymbol{p}_1=\boldsymbol{p}_1(\boldsymbol{P},\boldsymbol{p})$ and $\boldsymbol{p}_2=\boldsymbol{p}_2(\boldsymbol{P},\boldsymbol{p})$. 

The spins $s_1$ and $s_2$ of electron and positron can add up either to  a total spin $S=0$ (para-Ps) or to $S=1$ (ortho-Ps). In the case of para-Ps we consider the spin wave function~\cite{greiner} 
\begin{equation}
  \mathcal{R}_{s_1,s_2}^{0,0}=
  \frac{1}{\sqrt{2}}\left(\updelta_{{s_1},+}\updelta_{{s_2},-}-
\updelta_{{s_1},-} \updelta_{{s_2},+}\right)
  \label{eq:app_R_s1s2}
\end{equation}
which forms a singlet with only one possible configuration ($S=0,S_z=0$).   

\begin{figure}
    \centering
    \includegraphics{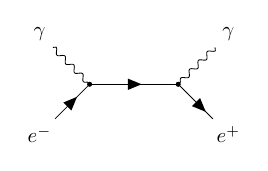}
    \includegraphics{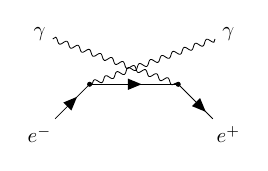}
    \caption{The two  Feynman diagrams corresponding to pair annihilation of electron ($e^-$) and positron ($e^+$) to two photons ($\gamma$). The direction of time is from bottom to top.}
    \label{fig:Ps}
\end{figure}

Applying the  terms of the $\hat{S}$-operator  in second-order perturbation theory~\cite{greiner} that are relevant for annihilation  yields the expressions
\begin{equation}
   \begin{aligned}
     \ket{\psi_\text{p-Ps}}&\propto \int\!\!\D^3\boldsymbol{k}_1\,\D^3\boldsymbol{k}_2\,\D^3\boldsymbol{P}\,\D^3\boldsymbol{p}\, \Psi(\boldsymbol{P})\, \psi(\boldsymbol{p}) \\
   &\times \, \updelta^4(p_1+p_2-k_1-k_2) \ket{\psi_\text{p-Ps}(\boldsymbol{k}_1,\boldsymbol{k}_2,\boldsymbol{p}_1,\boldsymbol{p}_2)}
   \end{aligned}
   \label{eq:app_psi_para_allg}
\end{equation}
and
\begin{equation}
 \begin{aligned}
 \ket{\psi_\text{p-Ps}(\boldsymbol{k}_1,\boldsymbol{k}_2,\boldsymbol{p}_1,\boldsymbol{p}_2)}&=\sum\limits_{j,l}
 \sum\limits_{s_1,s_2} \mathcal{R}_{s_1,s_2}^{0,0}
 \\ \times \, M_{s_1,s_2}^{j,l}(\boldsymbol{k}_1,\boldsymbol{k}_2,&\boldsymbol{p}_1,\boldsymbol{p}_2)\,
 \hat{a}_j^\dagger(\boldsymbol{k}_1)\,\hat{a}_l^\dagger(\boldsymbol{k}_2)\ket{0}
   \end{aligned}
   \label{eq:app_psi_para_k_allg}
\end{equation}
describing the emission of two photons with wave vectors $\boldsymbol{k}_1$ and $\boldsymbol{k}_2$.
The matrix element
\begin{equation}
 \begin{aligned}
M_{s_1,s_2}^{j,l}(\boldsymbol{k}_1,\boldsymbol{k}_2,\boldsymbol{p}_1\,\boldsymbol{p}_2)=
\bar{v}_{s_2}(\boldsymbol{p}_2)\,
\Gamma^{j,l}(\boldsymbol{k}_1,\boldsymbol{k}_2,\boldsymbol{p}_1)\, u_{s_1}(\boldsymbol{p}_1)
   \end{aligned}
   \label{eq:app_M_para_allg}
\end{equation}
depends on the elementary Dirac spinors $u_{s_1}=u_{s_1}(\boldsymbol{p}_1)$ and $\bar{v}_{s_2}=\bar{v}_{s_2}(\boldsymbol{p}_2)$ while the term
\begin{equation}
\begin{aligned}
\Gamma^{j,l}(\boldsymbol{k}_1,\boldsymbol{k}_2,\boldsymbol{p}_1)
=
\slashed{\varepsilon}_j^*(\boldsymbol{k}_1)\frac{1}{\slashed{p}_1-\slashed{k}_2-m}\slashed{\varepsilon}_l^*(\boldsymbol{k}_2)\,\\ +\,\slashed{\varepsilon}_l^*(\boldsymbol{k}_2)\frac{1}{\slashed{p}_1-\slashed{k}_1-m}\slashed{\varepsilon}_j^*(\boldsymbol{k}_1)
   \end{aligned}
\label{eq:app_para_Gamma}
\end{equation}
contains the four momenta $p_1,p_2,k_1,k_2$ of the particles and the polarization four vectors $\varepsilon_j^*{}^\mu=(0,\boldsymbol{\varepsilon}_j^*(\boldsymbol{k}_1))$, and $\varepsilon_l^*{}^\mu=(0,\boldsymbol{\varepsilon}_l^*(\boldsymbol{k}_2))$ of the photons.  We use the notation $\slashed{a}\equiv \gamma^\mu a_\mu=\gamma_\mu a^\mu$ with the Dirac matrices $\gamma^\mu$ and a four-vector $a$. In Fig.~\ref{fig:Ps} we have depicted the two Feynman diagrams that correspond to the matrix element given in Eqs.~\eqref{eq:app_M_para_allg} and~\eqref{eq:app_para_Gamma}.   

We assume that the Ps is at rest and the non-relativistic motion of its constituents can be neglected~\cite{sen2019}, that is $p_{1}^{\mu}=p_{2}^{\mu}\cong (m,0,0,0)$. In this approximation the center-of-mass momentum
$\boldsymbol{P}=\boldsymbol{p}_1+\boldsymbol{p}_2$
and the relative momentum
$\boldsymbol{p}=(\boldsymbol{p}_1-\boldsymbol{p}_2)/2$
are also zero, that is $\boldsymbol{P}=\boldsymbol{p}\cong0$.
If the center-of-mass wave function can be approximated by a delta function, $\Psi(\boldsymbol{P})\propto \updelta^3(\boldsymbol{P})$, the integration with respect to $\boldsymbol{P}$ can be easily performed. Similarly, the  ground-state wave function for the relative momentum varies on a typical scale of $\sim(\alpha/2)mc \ll mc $~ with $\alpha$ denoting  the fine-structure constant~\cite{itzykson}. Consequently, the integration over $\boldsymbol{p}$ yields the wave function in position space at the center of the Ps, that is $\psi(\boldsymbol{x}=0)$,  that is only a constant factor. Due to the delta function in Eq.~\eqref{eq:app_psi_para_allg} the wave vectors of the photons have opposite directions with  $\boldsymbol{k}_1= -\boldsymbol{k}_2\equiv \boldsymbol{k}$ but the same modulus $k=m$ which means an energy equal to the rest mass of an electron. We note that the corrections to this non-relativistic approximation scale with powers of $\boldsymbol{P}/m$ and  $\boldsymbol{p}/m$, respectively.  

With the help of the elementary relation $(\slashed{p}-m)u(\boldsymbol{p})=0$~\cite{bjorken} we obtain the expression~\cite{sen2019}
\begin{equation}
  \Gamma^{j,l}(\boldsymbol{k},-\boldsymbol{k})
=\frac{1}{2m^2} \left[\slashed{\varepsilon}_j^*(\boldsymbol{k})\slashed{\bar{k}} \slashed{\varepsilon}_l^*(-\boldsymbol{k}) +
\slashed{\varepsilon}_l^*(-\boldsymbol{k})\slashed{k} \slashed{\varepsilon}_j^*(\boldsymbol{k})\right]\,,
\label{eq:app_Gamma_nr}
\end{equation}
where we have defined $\bar{k}^\mu\equiv(k,-\boldsymbol{k})$.
%
We proceed by rewriting the spin wave function as 
$\displaystyle\mathcal{R}_{s_1,s_2}^{0,0}=2^{-1/2}\sum\limits_s s \, \updelta_{s,s_1}\updelta_{s,-s_2}$. By employing the identity $\gamma^5 u_s=s v_{-s}$~\cite{jauch} and the spin projection operator $\displaystyle (1+\gamma^0)/2 $~\cite{bjorken} for $p=0$ we arrive at~\cite{jauch}
\begin{equation}
    \sum_{s_1,s_2}\mathcal{R}_{s_1,s_2}^{0,0}M_{s_1,s_2}^{j,l}(\boldsymbol{k},-\boldsymbol{k})=\frac{1}{2\sqrt{2}}\trace{\gamma^5 \Gamma^{j,l}(\boldsymbol{k},-\boldsymbol{k})\gamma^0}
\label{eq:app_para_trace} 
\end{equation}
for the spin summation in Eq.~\eqref{eq:app_psi_para_k_allg}, where we have already used that the trace of the $\gamma^5$-matrix times an odd number of $\gamma$-matrices vanish~\cite{bjorken}.

Finally, we obtain the state 
\begin{equation}
    \ket{\psi_{\text{p-Ps}}}\propto \int\!\! \D^3 k \,
    \updelta(k-m)\ket{\psi_{\text{p-Ps}}(\boldsymbol{k})} 
\label{eq:app_psi_para}
\end{equation}
of the two emitted photons with 
\begin{equation}
  \ket{\psi_{\text{p-Ps}}(\boldsymbol{k})}=\frac{1}{\sqrt{2}}
  \sum\limits_{j,l} \mathcal{M}_{j,l}(\boldsymbol{k})
  \ket{j(\boldsymbol{k}),l(-\boldsymbol{k})}\,
  \label{eq:app_psi_para_k}
\end{equation}
and the matrix element
\begin{equation}
    \mathcal{M}_{j,l}(\boldsymbol{k})=i\left[\boldsymbol{\varepsilon}_j^*(\boldsymbol{k})\times \boldsymbol{\varepsilon}_l^*(\boldsymbol{-k})\right]\cdot \hat{\boldsymbol{k}}=j\,\updelta_{j,l}\,,
    \label{eq:app_M_para}
\end{equation}
where we have introduced the unit vector $\hat{\boldsymbol{k}}\equiv \boldsymbol{k}/k$ in $\boldsymbol{k}$-direction. To find the first expression, we have evaluated the trace in Eq.~\eqref{eq:app_para_trace} with the help of elementary identities~\cite{bjorken}. For the second one, we have assumed that both photons are circularly polarized, that is $\boldsymbol{\varepsilon}_{\pm}(\boldsymbol{k})= (\hat{\boldsymbol{x}}_\text{A} \pm i \hat{\boldsymbol{y}}_\text{A})/\sqrt{2}$  and $\boldsymbol{\varepsilon}_{\pm}(-\boldsymbol{k})= (\hat{\boldsymbol{x}}_\text{B} \pm i  \hat{\boldsymbol{y}}_\text{B})/\sqrt{2}$, and  that the linear polarization vectors in $y$-direction coincide $\hat{\boldsymbol{y}}_\text{A}=\hat{\boldsymbol{y}}_\text{B}$ while $\hat{\boldsymbol{x}}_\text{A}=-\hat{\boldsymbol{x}}_\text{B}$ (compare to Fig.~\ref{fig:setup}). We note that Eqs.~\eqref{eq:app_psi_para},~\eqref{eq:app_psi_para_k}  and~\eqref{eq:app_M_para}  give rise to Eqs.~\eqref{eq:psi_para} and~\eqref{eq:psi_k_para} in the main part of this paper.    

\section{Compton scattering}
\label{sec:Compton_scattering}
In this appendix, we derive the angular distributions for photons from Compton scattering. Instead of employing the Stokes vector formalism~\cite{fano1949,mcmaster1961}, we directly work with the density matrix~\cite{wightman1948}. We generalize the situation for a single photon to coincidences of many photons and finally apply these results to the two-photon case relevant for the para-Ps decay.      

\subsection{One photon}
We start with the simplest case, that is the scattering of one photon from one electron. The initial state of this system is given by the product
\begin{equation}
 \hat{\rho}_{N=1}^{\text{in}}= \sum\limits_{j,l}\rho_{j,l} \ket{j(\boldsymbol{k})} 
 \bra{l(\boldsymbol{k})}\otimes \frac{1}{2}\sum\limits_s 
 \ket{s(\boldsymbol{p})} \bra{s(\boldsymbol{p})}\,,
\end{equation}
where $\rho_{j,l}$ denotes the density matrix of the photon with respect to its polarization while the electron is completely unpolarized. We assume that both, photon and electron, are described by plane waves with definite wave vector $\boldsymbol{k}$ and momentum $\boldsymbol{p}$, respectively. Moreover, the electron is initially at rest, that is $\boldsymbol{p}=0$. For convenience, we use a notation, where $\ket{j(\boldsymbol{k})}\equiv\hat{a}_j^\dagger(\boldsymbol{k})\ket{0}$ and $\ket{s(\boldsymbol{p})}\equiv\hat{b}_s^\dagger(\boldsymbol{p})\ket{0}$.

\begin{figure}
    \centering
    \includegraphics{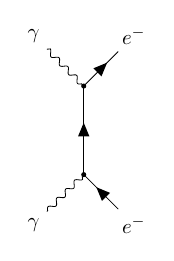}
    \includegraphics{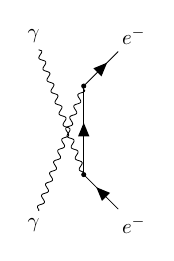}
    \caption{The two  Feynman diagrams corresponding to Compton scattering of a photon ($\gamma$) with an electron ($e^-$). The direction of time is from bottom to top.}
    \label{fig:Compton}
\end{figure}

After scattering, the density operator of the total system reads
\begin{equation}
\hat{\rho}_{N=1}^\text{out}\propto
\frac{1}{2}
\sum\limits_{j,l}\sum\limits_s
\rho_{j,l}
\left[\hat{S}\ket{j(\boldsymbol{k}),s(\boldsymbol{p})}\right]
\left[\hat{S}\ket{l(\boldsymbol{k}),s(\boldsymbol{p})}\right]^\dagger \,,
\label{eq:app_rho_out}
\end{equation}
where the $\hat{S}$-operator in second-order perturbation theory acts on the basis states yielding
\begin{equation}
 \begin{aligned}
  \hat{S}\ket{j(\boldsymbol{k}),s(\boldsymbol{p})}\propto
  \!\frac{m}{\sqrt{E_{\boldsymbol{p}}|\boldsymbol{k}|}}
 \!\int\!\!\!\frac{\text{d}^3p'}{\sqrt{E_{\boldsymbol{p}'}}}
  \int\!\!\!\frac{\text{d}^3k'}{\sqrt{|\boldsymbol{k}'|}}
   \,\updelta(p'+k'-p-k)\\
  \times \sum\limits_{s',m}
  M_{s,s'}^{j,m}(\boldsymbol{k},\boldsymbol{k}',\boldsymbol{p},\boldsymbol{p}')
  \ket{m(\boldsymbol{k'}),s'( \boldsymbol{p'})}
 \end{aligned}
 \label{eq:app_Scomp_psi}
\end{equation}
with the matrix element $M_{s,s'}^{j,m}$ given by the relations
\begin{equation}
   M_{s,s'}^{j,m}(\boldsymbol{k},\boldsymbol{k}',\boldsymbol{p},\boldsymbol{p}') \equiv
   (-ie)^2 \bar{u}_{s'}(\boldsymbol{p}')\,\Gamma_\text{C}^{j,m}(\boldsymbol{k},\boldsymbol{k}',\boldsymbol{p})\,u_s(\boldsymbol{p})
   \label{eq:app_Mcomp}
\end{equation}
and
\begin{equation}
 \begin{aligned}
       \Gamma_\text{C}^{j,m}(\boldsymbol{k},\boldsymbol{k}',\boldsymbol{p})\equiv \slashed{\varepsilon}_m^*(\boldsymbol{k}')\frac{1}{\slashed{p}+\slashed{k}-m}\slashed{\varepsilon}_j(\boldsymbol{k})
\\
+\slashed{\varepsilon}_j(\boldsymbol{k})\frac{1}{\slashed{p}-\slashed{k}'-m}\slashed{\varepsilon}_m^*(\boldsymbol{k}')\,.
 \end{aligned}
\end{equation}
These expressions correspond to the two Feynman diagrams of Compton scattering depicted in Fig.~\ref{fig:Compton}.

If we measure only the degrees of freedom of the photon, we restrict ourselves to the reduced density operator 
\begin{equation}
    \hat{\rho}_{N=1}^\text{phot}=\sum\limits_{\sigma}\int\!\!\D^3\wp \,\hat{b}_\sigma(\boldsymbol{\wp})\hat{\rho}_{N=1}^\text{out}\hat{b}_\sigma^\dagger(\boldsymbol{\wp})
    \label{eq:app_rho_1phot}
\end{equation}
for the photon, that is, we average over the subsystem of the electron. Ultimately, we measure the angular distribution of the scattered photons, that are described by the wave vector $\boldsymbol{\varkappa}=\boldsymbol{\varkappa}(\varkappa,\theta,\phi)$. This distribution is given by the expression
\begin{equation}
 p(\theta,\phi)=
 \int\!\!\D\varkappa \, \varkappa^2 
  \sum\limits_\lambda \trace{\hat{a}_\lambda(\boldsymbol{\varkappa})\hat{\rho}_{N=1}^\text{phot}\hat{a}_\lambda^\dagger(\boldsymbol{\varkappa})}\,,
  \label{eq:app_p_allg_1phot}
\end{equation}
where we integrate over the radial degrees of freedom. Due to the delta function in Eq.~\eqref{eq:app_Scomp_psi} the wave number of the scattered photon depends via the Compton formula
\begin{equation}
    \varkappa(\theta)=\frac{k}{1+\frac{k}{m}(1-\cos{\theta})}
\end{equation}
on the scattering angle $\theta$, that is, the angle between $\boldsymbol{k}$ and $\boldsymbol{\varkappa}$. With the help of Eqs.~\eqref{eq:app_rho_out} --%
~\eqref{eq:app_p_allg_1phot} we find 
\begin{equation}
   p(\theta,\phi)=\mathcal{N} \left(\frac{\varkappa(\theta)}{k}\right)^2
   \trace{\mathcal{V}(\theta,\phi)\cdot \rho}
   \label{eq:app_p_1phot}
\end{equation}
with the elements
\begin{equation}
 \begin{aligned}
    \mathcal{V}_{j,l}(\theta,\phi)\!&\equiv 
  \!\!\!\sum\limits_{\lambda,s,\sigma}
  \!\!\! M_{s\sigma}^{j\lambda}(\boldsymbol{k},\boldsymbol{\varkappa},0,\boldsymbol{k}\!-\!\boldsymbol{\varkappa})\!\left[M_{s\sigma}^{l\lambda}(\boldsymbol{k},\boldsymbol{\varkappa},0,\boldsymbol{k}\!-\!\boldsymbol{\varkappa})\right]^\dagger
  \\
  =\sum\limits_\lambda &\trace{\frac{\slashed{\wp}+m}{2m}\Gamma_\text{C}^{j,\lambda}(\boldsymbol{k},\boldsymbol{\varkappa},\boldsymbol{p})\frac{\slashed{p}+m}{2m}\bar{\Gamma}_\text{C}^{l,\lambda}(\boldsymbol{k},\boldsymbol{\varkappa},\boldsymbol{p})}
 \end{aligned}
\end{equation}
of the matrix $\mathcal{V}=\mathcal{V}(\theta,\phi)$. Here we have omitted the explicit dependency of $\boldsymbol{\varkappa}$ on the angles, that is $\boldsymbol{\varkappa}=\boldsymbol{\varkappa}(\theta,\phi)$. Moreover, we note the identities
$\boldsymbol{\wp}=\boldsymbol{k}-\boldsymbol{\varkappa}$, 
$\boldsymbol{p}=0$, and the definition
$\bar{\Gamma}_\text{C}\equiv \gamma^0 \Gamma_\text{C}^\dagger \gamma^0$. The constant $\mathcal{N}$ is determined
by normalizing the distribution to unity with respect to the entire solid angle of $4\pi$.  

\begin{figure}
    \centering
    \includegraphics[width=0.9\columnwidth]{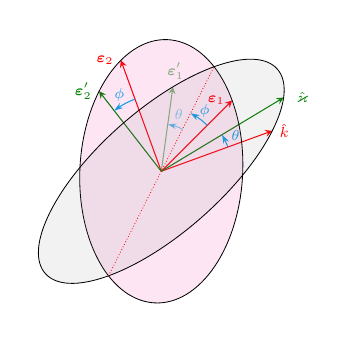}
    \caption{Geometry of Compton scattering in terms of a linear polarization basis: The initial set of basis vectors, $\boldsymbol{\varepsilon}_1$, $\boldsymbol{\varepsilon}_2$,  $\hat{\boldsymbol{k}}$  and the final set $\boldsymbol{\varepsilon}'_1$,$\boldsymbol{\varepsilon}'_2$, $\hat{\boldsymbol{\varkappa}}$ are connected by the relations Eqs.~\eqref{eq:epsilon_1},~\eqref{eq:epsilon_2}, and~\eqref{eq:n_k}~\cite{jackson}, where $\theta$ represents the scattering angle while $\phi$ is the azimuthal angle.}
    \label{fig:kosy}
\end{figure}

After a tedious calculation, we obtain~\cite{wightman1948}
\begin{equation}
 \begin{aligned}
      \mathcal{V}_{j,l}(\theta,\phi)=&
  \left(\frac{\varkappa(\theta)}{k}+\frac{k}{\varkappa(\theta)}-2\right) \updelta_{j,l} \\
  &+2\sum\limits_{\lambda} 
  \boldsymbol{\varepsilon}_j(\boldsymbol{k})\cdot \boldsymbol{\varepsilon}'_\lambda(\boldsymbol{\varkappa})
  \,  \boldsymbol{\varepsilon}_l(\boldsymbol{k})\cdot \boldsymbol{\varepsilon}'_\lambda(\boldsymbol{\varkappa})\,,
 \end{aligned}
 \label{eq:V_scalar_pr}
\end{equation}
where we have employed fundamental trace identities for $\gamma$-matrices and have expanded both, initial and scattered, fields in linear polarization-bases with real polarization vectors.
We choose the geometry such that the final  set of basis vectors, $\boldsymbol{\varepsilon}'_1$,$\boldsymbol{\varepsilon}'_2$, $\hat{\boldsymbol{\varkappa}}$, and the initial set $\boldsymbol{\varepsilon}_1$,$\boldsymbol{\varepsilon}_2$,  $\hat{\boldsymbol{k}}$ are connected by the relations~\cite{jackson}
\begin{align}
\label{eq:epsilon_1}
\boldsymbol{\varepsilon}'_1&=\cos{\theta}\left(\boldsymbol{\varepsilon}_1\cos\phi+\boldsymbol{\varepsilon}_2\sin\phi\right)-\hat{\boldsymbol{k}}\sin\theta\, ,\\
\label{eq:epsilon_2}
\boldsymbol{\varepsilon}'_2&=- \boldsymbol{\varepsilon}_1\sin{\phi}
+\boldsymbol{\varepsilon}_2\cos{\phi} \, ,\\
\label{eq:n_k}
\hat{\boldsymbol{\varkappa}}&=\sin{\theta}
\left(\boldsymbol{\varepsilon}_1 \cos{\phi}+
\boldsymbol{\varepsilon}_2 \sin{\phi}\right)
+ \hat{\boldsymbol{k}}\cos{\theta}
\end{align}
defining the scattering angle $\theta$ and azimuth angle $\phi$ with $0\leq \theta < \pi$ and $0\leq \phi < 2\pi$, respectively.   This geometry, illustrated in Fig.~\ref{fig:kosy}, leads finally to the expression
\begin{equation}
  \mathcal{V}(\theta,\phi)=
  S(\theta)\,\mathds{1}-f(\theta)
   \left[\cos{(2\phi)} \,\sigma_z +\sin{(2\phi)}\,\sigma_x \right]
   \label{eq:V}
\end{equation}
for the matrix $\mathcal{V}$ with the definitions from Eq.~\eqref{eq:S_theta} and~\eqref{eq:f_theta}.

We can rewrite the distribution in Eq.~\eqref{eq:app_p_1phot} as 
\begin{equation}
    p(\theta,\phi)=\mathcal{N} \,w^\text{T}\cdot v(\theta,\phi)
\end{equation}
with the help of Eqs.~\eqref{eq:V} and~\eqref{eq:v} and by defining the elements $w_j\equiv \trace{u_j\cdot \rho}$  of a column vector $w$ with $u$ from Eq.~\eqref{eq:u}.  

For a photon initially linearly polarized in $x$-direction, that is $\rho_{j,l}=\updelta_{j,1}\updelta_{l,1}$ we straightforwardly find a 
distribution $p\propto \mathcal{V}_{11}$
which is of course identical to the Klein--Nishina cross section for polarized photons. Similarly, we obtain for completely unpolarized photons with $\displaystyle\rho=\frac{1}{2}\mathds{1}$ a distribution $p\propto S(\theta)$ that is independent of $\phi$.

\subsection{Generalization to many photons}

Now, we generalize our results of the preceding section to a situation with $N$ modes each containing a single photon. For that purpose, we assume a local interaction of each photon with one certain electron due to the macroscopic distances $\sim d$ between the scatterers. 
The width $\Delta r$  of the electron wave packets has to be in the regime $k^{-1} \ll \Delta r  \ll d$, such that (i) the interaction is localized and (ii) the electron wave functions can be approximated as plane waves ($k^{-1}\sim 10^{-12}$\,m for annihilation photons).

The final density operator for the total system reads
\begin{equation}
 \begin{aligned}
 \hat{\rho}_{N}^\text{out}  \propto 
2^{-N}&\!\!\!\!\!\!\!\!\!\sum\limits_{j_1,l_1,...,j_N,l_N}\,\sum\limits_{s_1,...,s_N}\rho_{j_1,l_1,...,j_N,l_N}\\
\bigotimes\limits_{n=1}^N&
\left[\hat{S}\ket{j_n(\boldsymbol{k}_n),s_n(\boldsymbol{p}_n)}\right]
\left[\hat{S}\ket{l_n(\boldsymbol{k}_n),s_n(\boldsymbol{p}_n)}\right]^\dagger
 \end{aligned}
 \label{eq:app_rho_out_N}
\end{equation}
which reduces to the initial state when we let $\hat{S} \rightarrow \mathds{1}$. The density operator for the photons is given by the expression
\begin{equation}
  \begin{aligned}
    \hat{\rho}_{N}^\text{phot}&=
    \sum\limits_{\sigma_1,...,\sigma_N} 
    \int\!\!\D^3\wp_1 \, ... \int\!\!\D^3\wp_N 
    \\
    &\times \,  \hat{B}_{\sigma_1,...,\sigma_N}(\boldsymbol{\wp}_1,...,\boldsymbol{\wp}_N)\hat{\rho}_{N}^\text{out}
    \hat{B}_{\sigma_1,...,\sigma_N}^\dagger(\boldsymbol{\wp}_1,...,\boldsymbol{\wp}_N)\,,
  \end{aligned}
\end{equation}
where we have introduced the operator
\begin{equation}
   \hat{B}_{\sigma_1,...,\sigma_N}(\boldsymbol{\wp}_1,...,\boldsymbol{\wp}_N) \equiv \bigotimes\limits_{n=1}^N
   \hat{b}_{\sigma_n}(\boldsymbol{\wp}_n)
\end{equation}
for the electrons. We then calculate the the probability for photon coincidences via the relation
\begin{equation}
\begin{aligned}
      p(\theta_1,\phi_1,..., \theta_N,\phi_N)=\sum\limits_{\lambda_1,...,\lambda_N}
    \int\!\!\D\varkappa_1 \varkappa_1^2 \,...
    \int\!\!\D\varkappa_N \varkappa_N^2
     \\
    \times\trace{\hat{A}_{\lambda_1,...,\lambda_N}(\boldsymbol{\varkappa}_1,...\boldsymbol{\varkappa}_N)
    \hat{\rho}_{N}^\text{phot}\hat{A}_{\lambda_1,...,\lambda_N}^\dagger(\boldsymbol{\varkappa}_1,...\boldsymbol{\varkappa}_N)}
\end{aligned}
\end{equation}
after introducing the operator
\begin{equation}
   \hat{A}_{\lambda_1,...,\lambda_N}(\boldsymbol{\varkappa}_1,...,\boldsymbol{\varkappa}_N) \equiv \bigotimes\limits_{n=1}^N
   \hat{a}_{\lambda_n}(\boldsymbol{\varkappa}_n)
   \label{eq:app_A_N}
\end{equation}
for the photons. It is easy to see that the prescription in Eqs.~\eqref{eq:app_rho_out_N} to~\eqref{eq:app_A_N}  consists of repeating the procedure in the preceding section for a single photon $N$ times.  Finally, we arrive at the result
\begin{equation}
 \begin{aligned}
 p&(\theta_1,\phi_1,...\theta_N,\phi_N)\propto
\prod_{n=1}^N\left(\frac{\varkappa(\theta_n)}{k_n}\right)^2
 \sum\limits_{j_1,l_1,...,j_N,l_N}\\
&  \mathcal{V}_{j_1,l_1}(\theta_1,\phi_1)...
\mathcal{V}_{j_N,l_N}(\theta_N,\phi_N)
\, \rho_{j_1,l_1,...,j_N,l_N}\,,
 \end{aligned}
 \label{eq:p_allg}
\end{equation}
where the contribution of each mode is weighted with the density matrix $\rho_{j_1,l_1,...,j_N,l_N}$ for $N$ photons.

\subsection{Two photons}

We illustrate the procedure for calculating coincidences from Compton scattering for the example of two emitted photons from the p-Ps decay. For that purpose, we consider the initial state $\ket{\psi}=\ket{\psi_\text{p-Ps}(\boldsymbol{k})}$ from Eq.~\eqref{eq:psi_k_para}. 
By setting $N=2$ in Eq.~\eqref{eq:p_allg} we obtain the relation
\begin{equation}
 \begin{aligned}
     p(\theta_\text{A},\phi_\text{A},\theta_\text{B},\phi_\text{B})=\mathcal{N} 
    \left(\frac{\varkappa(\theta_\text{A})}{m}\right)^2\left(\frac{\varkappa(\theta_\text{B})}{m}\right)^2\\
    \times\trace{\mathcal{V}(\theta_\text{A},\phi_\text{A})\otimes \mathcal{V}(\theta_\text{B},\phi_\text{B})\cdot \, \rho}\,,
 \end{aligned}
\end{equation}
where we have changed the subscripts $1,2 \rightarrow \text{A},\text{B}$ to connect with the notation in the main part of the article. Inserting the explicit representation of $\mathcal{V}$ from Eq.~\eqref{eq:V} leads to Eq.~\eqref{eq:p} by means of Eqs.~\eqref{eq:v},~\eqref{eq:W}, and~\eqref{eq:u}. With the help of Eq.~\eqref{eq:rho_psi_+} we finally recover the expression from Eq.~\eqref{eq:p_Ps}
for the probability of coincidences from the para-Ps decay. 

\begin{figure}
    \centering
    \includegraphics[width=0.85\columnwidth]{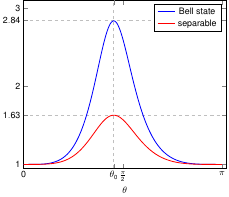}
    \caption{Ratio of counts for perpendicular and (anti-)parallel detection of coincidences for a Bell state (blue curve), Eq.~\eqref{eq:r}, and the constructed separable state (red curve),  Eq.~\eqref{eq:r_sep},  respectively. We study the dependency on the scattering angle $\theta_\text{A}=\theta_\text{B}\equiv \theta$. In both cases, this ratio attains its maximum at $\theta_0\cong 81.7^\circ$, but the maximum value for the separable state is with $1.63$ smaller compared to the Bell state with $2.84$~\cite{bohm1957,caradonna2024}.}
    \label{fig:sep}
\end{figure}

In the following we review the arguments of Refs.~\cite{bohm1957,caradonna2024}, where a hypothetical separable state was constructed. 
The density operator~\cite{bohm1957}
\begin{equation}
  \hat{\rho}_\text{sep}\!=\!\frac{1}{2\pi} \!\int\limits_0^{2\pi}\!\!\D\alpha \left[D(-\alpha)\ket{1}
  D(\alpha)\ket{2} \right]\!
  \left[\bra{1}D^\dagger(-\alpha)
  \bra{2} D^\dagger(\alpha) \right]
\end{equation}
describes a mixture of perpendicularly polarized photons which are rotated with the help of the rotation matrix~\cite{caradonna2024}
\begin{equation}\
    D(\alpha)\equiv
    \begin{pmatrix}
    \cos{\alpha} & \sin{\alpha} \\
     -\sin{\alpha} & \cos{\alpha}
    \end{pmatrix}
\end{equation}
by the angles $~\pm \alpha$. This construction ensures that the average angular momentum is conserved~\cite{bohm1957}. We obtain the expression
\begin{equation}
    \rho_\text{sep}=\frac{1}{8}
    \begin{pmatrix}
    1 & 0 & 0 & 1\\
    0 & 3 & 1 & 0\\
    0 & 1 & 3 & 0\\
    1 & 0 & 0 & 1
    \end{pmatrix}
\end{equation}
that leads to the distribution
\begin{equation}
      p(\theta_\text{A},\!\phi_\text{A},\!\theta_\text{B},\!\phi_\text{B})\!\propto\! 
      S(\theta_\text{A})S(\theta_\text{B})-\frac{1}{2}f(\theta_\text{A})\!f(\theta_\text{B})\!\cos{\!\left[2(\!\phi_\text{A}\!+\!\phi_\text{B}\!)\!\right]}
\end{equation}
of coincidences by means of Eq.~\eqref{eq:p}. At first sight this result looks similar to the distribution for the Bell state in Eq.~\eqref{eq:p_Ps}. However, the pre-factor in front of the cosine is $1/2$ instead of $1$. Hence, the ratio in Eq.~\eqref{eq:r} would lead to 
\begin{equation}
    \frac{p(\theta_0 | 0, \pi/2)}{p(\theta_0|0,0)}=
    \frac{1+a^2(\theta_0)/2}{1-a^2(\theta_0)/2}\cong 1.63
    \label{eq:r_sep}
\end{equation}
instead of $2.84$. This different behavior is illustrated in Fig.~\ref{fig:sep}. 

We note that the authors of Ref.~\cite{bohm1957} considered also a product state of two circularly polarized photons $\ket{+,+}$ or $\ket{-,-}$. These states are already symmetric with respect to a rotation around the $z$-axis, that is  $D(-\alpha)\ket{\pm} D(\alpha)\ket{\pm}=\ket{\pm,\pm}$, and an integration over $\alpha$ would not have any effect. For these input states the resulting distribution 
$p(\theta_\text{A},\phi_\text{A},\theta_\text{B},\phi_\text{B})\!\propto\! 
      S(\theta_\text{A}\!)S(\theta_\text{B})$
of coincidences is independent of the azimuths which can be immediately ruled out by the experimental results~\cite{wu1950}.

\subsection{Multiple Compton scattering events}

In the following, we briefly discuss the setup from Refs.~\cite{caradonna2024_annals,bordes2024}, where one of the two photons is scattered two times, that is on one side is an additional scatterer. We have to adapt our formalism to this situation and calculate the angular distribution of coincidences via the prescription 
\begin{equation}
\begin{aligned}
        p(\theta_\text{A},\phi_\text{A},\theta_{\text{B}2},\phi_{\text{B}2};\theta_{\text{B}1})\propto
    \sum\limits_{j_1,l_1,j_2,l_2}\rho_{j_1l_1,j_2,l_2}\mathcal{V}_{j_1l_1}(\theta_\text{A},\phi_\text{A})\\
    \times \sum\limits_{m,n} \mathcal{V}_{mn}(\theta_{\text{A}2},\phi_{\text{A}2};\theta_{\text{A}1})\mathcal{V}_{mn|j_2l_2}(\theta_{\text{B}1})\,,
\end{aligned}
\label{eq:last}
\end{equation}
where $\theta_\text{A},\theta_{\text{B}2}$ and $\phi_\text{A},\phi_{\text{B}2}$ are the scattering and azimuthal angles for the detected photons, respectively, while $\theta_{\text{B}1}$ gives the scattering angle for the intermediate scattering process on one side of the setup. We set the corresponding azimuth to zero, that is $\phi_{\text{B}1}=0$. The matrix element 
\begin{equation}
\begin{aligned}
\mathcal{V}_{m,n|j,l}(\theta_{\text{B}1})= 
2\boldsymbol{\varepsilon}_j \cdot {\boldsymbol{\varepsilon}_m}'\,
\boldsymbol{\varepsilon}_l\cdot \boldsymbol{\varepsilon}_n'+\frac{1}{2}\left(S(\theta_{\text{B}1})+f(\theta_{\text{B}1})-2\right)\\
\left[\boldsymbol{\varepsilon}_j \cdot \boldsymbol{\varepsilon}_l\,\boldsymbol{\varepsilon}_n'\cdot {\boldsymbol{\varepsilon}_m}'
+\left(\boldsymbol{\varepsilon}_j \times \boldsymbol{\varepsilon}_l\right) \cdot
\left(\boldsymbol{\varepsilon}_m' \times \boldsymbol{\varepsilon}_n'\right)
\ 
\right]
\end{aligned}
\end{equation}
describes this intermediate scattering process. It is analogously derived as the expression for $\mathcal{V}_{j,l}$ in Eq.~\eqref{eq:V_scalar_pr}, but here we do not sum over the polarizations $m,n$ of the scattered photons and we do not set $m=n\equiv \lambda$.  

Inserting the density matrix for $\ket{\Psi}_+$ from Eq.~\eqref{eq:rho_psi_+} into Eq.~\eqref{eq:last} yields the same results as derived in Eqs. (30), (31),
and (32) of Ref.~\cite{caradonna2024_annals} with the help of an approach using the Stokes vector.   

\end{appendix}

\bibliography{references}

\end{document}